%% file: draft1_1.tex
\documentclass[aps,prl,reprint,floatfix]{revtex4-2}
\usepackage{placeins}
\usepackage{amsmath, amssymb, amsfonts, amsbsy, amsthm, latexsym, graphicx,mathtools}
\usepackage{xcolor}
\usepackage[scr=rsfs]{mathalpha}
\usepackage{subcaption} 
\usepackage{dcolumn}
\usepackage{bm}

\newcommand{\jump}[1]{\ensuremath{[\![#1]\!]}}

\newcommand{\DZ}[1]{{\color{magenta}  #1}}

\input{macros}

\begin{document}

\preprint{APS/123-QED}

\title{Exact coherent structures as building blocks of turbulence on large domains}

\author{Dmitriy Zhigunov}
\email{dmitriy.zhigunov@ed.ac.uk}
\author{Jacob Page}%
 \email{jacob.page@ed.ac.uk}
\affiliation{%
 School of Mathematics and Maxwell Institute for Mathematical Sciences, University of Edinburgh,
Edinburgh, EH9 3FD, UK
}%

\date{\today}

\begin{abstract}

Exact unstable solutions of the Navier-Stokes equations are thought to underpin the dynamics of turbulence, but are usually computed in minimal computational domains.   
Here, we extend this dynamical systems approach to spatially extended turbulent flows featuring multiple interacting `substructures’, and show how new simple invariant solutions can be constructed by spatial tiling of exact solutions from small-box calculations. 
Candidate solutions are found via gradient-based optimization of a scalar loss function which targets autorecurrence in spatially-masked regions of the flow. 
We apply these ideas to a vertically-extended Kolmogorov flow, where we first identify large numbers of relative periodic orbits (RPOs) which are combinations of high-dissipation, small-box solutions with laminar patches. 
We then show that vertically-stacked combinations of pairs of distinct small-box RPOs can form robust guesses for dynamically-relevant \emph{two-tori} in the larger domain. 
Finally, we show how our optimization procedure can identify ‘turbulent’ trajectories which locally shadow a small-box RPO for multiple periods in a subdomain. 
These small-box combinations are possible as the flow spends prolonged periods in a regime where it can be effectively considered as a pair of weakly-coupled small-box systems, due to shielding effects associated with higher-dissipation flow structures. 

\end{abstract}

\maketitle



The dynamical systems view of turbulence considers a realization of a particular flow as a trajectory in a high-dimensional state space, pinballing between exact, unstable solutions of the Navier Stokes equations \cite{kawahara2012}. This picture has been realised experimentally in both 2D \cite{suri2017} and 3D \cite{crowley2022} flows.
Early applications of these ideas explained many aspects of the subcritical transition to turbulence in parallel shear flows \cite{faisst2003,wedin2004}, resulting in 
intensive focus on the identification and computation of exact solutions in fully-developed turbulence \cite{kawahara2001,gibson2009,chandler2013}.
The exact solutions, which are typically relative periodic orbits (RPOs) due to the presence of continuous symmetries in many of the canonical fluid systems, each contain a closed cycle of physical processes helping to sustain the fully turbulent state \cite{Waleffe1997, Hall2010}. 
Computing RPOs is notoriously challenging, particularly in more strongly turbulent systems, and has restricted much of the work to date to weak turbulence in `minimal' computational domains (e.g. see the review \cite{graham2021}). 
While recent approaches based on optimization have produced enough solutions to reconstruct turbulent statistics at modest Reynolds numbers \cite{page2024}, $Re$, there are to our knowledge no examples of exact solutions featuring multiple spatially localized substructures typically associated with turbulence in larger domains. 



There have been notable examples of spatially localized exact solutions obtained in extended domains for various parallel shear flows. 
This includes the identification of spanwise-localized roll/streak solutions (and associated homoclinic snaking) obtained by \cite{schneider2010} and streamwise-localized RPOs computed at transitional $Re$ in a pipe \cite{avila2013Onset}. 
Solutions which localize in the \emph{cross shear} direction have also been found in planar configurations \cite{deguchi2015,eckhardt2018} and appear to continue indefinitely up in $Re$. 
A slightly different perspective was adopted by \cite{Beaume2016}, who sought new large-box exact solutions (in a reduced system of equations) which are spanwise-spatially-modulated small-box states.
However, these `substructures' are all dominated by a single cross-shear lengthscale, while here we explicitly seek new solutions with multiple substructures. 
The method is inspired by 
the ideas introduced by Cvitanovi\'c and coauthors \cite{liang2022,cvitanovic2025} describing spatiotemporal chaos as a space-time tapestry composed from smaller solutions.



In this Letter we construct spatially extended solutions of the Navier-Stokes equations from combinations of small-box exact solutions. 
Our starting point is a large library of RPOs assembled in two-dimensional Kolmogorov flow on a $2\pi \times 2\pi$ domain \cite{Cleary2025} which we seek to combine in a larger $2\pi \times 4\pi$ box.
The new solutions include (i) exact RPOs which are spatially-tiled combinations of a small box RPO with the laminar solution; (ii) unstable two-tori which are constructed from pairs of non-trivial small-box RPOs\DZ{;} and (iii) turbulent trajectories which locally shadow a small box solution in a subregion of the flow domain. 
Computation of the new states is performed with a combination of gradient-based optimization and classical Newton-Krylov methods, and opens up a promising avenue to seek multiscale solutions in three-dimensional, wall bounded flows. 



\begin{figure}
\includegraphics[width=\linewidth]{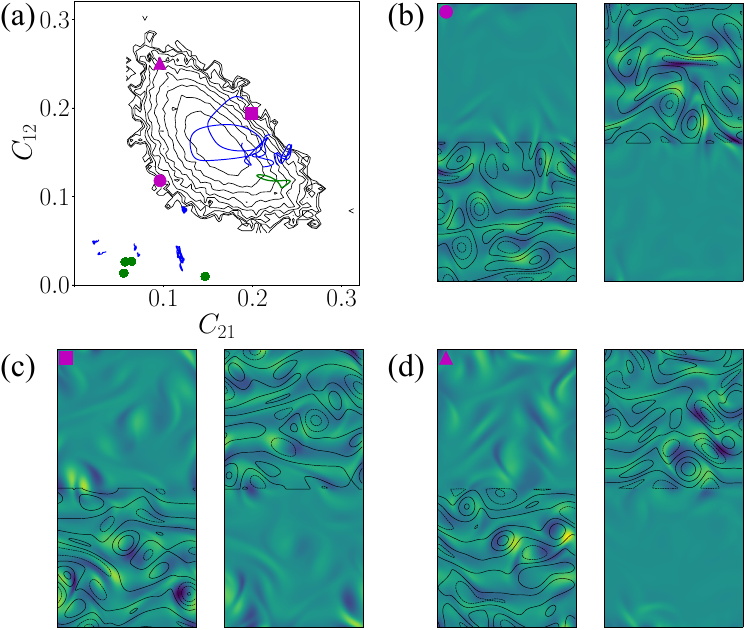}
\caption{
Cross-stream coupling in the tall-box Kolmogorov flow.
(a) Cross-coupling terms $C_{12}$ and $C_{21}$ (see equation \ref{eqn:coupling_defn}) for a turbulent $2\pi\times4\pi$ trajectory of total length $T = 10^5$ (contours; levels evenly spaced logarithmically from 0.4 to 228.0). Also shown are the coupling coefficients for the RPO-laminar tiles (green curves/points) and the RPO-RPO tiles (blue curves). (b, c, d) Examples of weakly coupled, two-way coupled and one-way coupled vorticity fields respectively, corresponding to the magenta symbols in (a). Background filled contours are the explicit coupling terms ${\bf u}_1\cdot\boldsymbol \nabla\omega$ (left) and ${\bf u}_2\cdot\boldsymbol \nabla\omega$ (right) with contours in ranges (a) $[-24.2,24.2]$, (b) $[-17.5,17.5]$, and (c) $[-25.0,25.0]$. Line contours show the vorticity, evenly spaced $-11.4<\omega<11.4$, in each half domain which is used to compute the induced velocities $\{\mathbf u_i\}$.
}
\label{fig:coupling} 
\end{figure}


We consider two-dimensional Kolmogorov flow driven by a monochromatic force in the horizontal, $\mathbf f^* := \hat{\mathbf x} \, \chi^* \sin(2 \pi n y^* / L^*)$ (an asterisk indicates a dimensional variable). 
The forcing amplitude, $\chi^*$, and fundamental wavenumber associated with the box width, $k^* := 2\pi / L^*$, are used to non-dimensionalize the system.
The out-of-plane vorticity evolves according to 
\begin{equation}
    \partial_t\omega + \mathbf u \cdot \boldsymbol\nabla\omega +\frac{1}{Re}\Delta \omega - n\cos(ny),
    \label{eqn:vort}
\end{equation}
where the velocity and vorticity are related by the definition $\omega := (\boldsymbol \nabla \times \mathbf u)\cdot \hat{\mathbf z}$, while the Reynolds number based on our non-dimensionalization is $Re := (1/\nu) \sqrt{\chi^* / k^{*3}}$.
We fix the number of forcing waves per box \emph{width} at $n=4$ but contrary to previous studies \cite{chandler2013,page2024} consider `tall' boxes of size $(L_x, L_y) = (2\pi, 4\pi)$ (so that there are eight forcing waves in the box). The Reynolds number is held fixed at $Re =40$ throughout this paper. 
Equation (\ref{eqn:vort}) is equivariant under continuous translations in $x$, under discrete shift-reflects over half-wavelengths of the forcing in $y$ and under rotation through $\pi$ \cite{chandler2013}.

We seek relative periodic/quasi-periodic solutions in the $4\pi \times 2\pi$ box which are constructed from RPOs found in the `standard' $2\pi\times 2\pi$ geometry. 
Our expectation that this is possible is due to the fact that the tall box can often be considered as a pair of weakly interacting small-boxes. 
To quantify the interaction, consider a partitioning of the vorticity in the vertical coordinate, $\omega(x,y)=\omega(x,y)\,\mathbf{1}_{[0,2\pi)}(y)+\omega(x,y)\,\mathbf{1}_{[2\pi,4\pi)}(y):= \omega_1 + \omega_2$, with matching conditions across the interfaces $\jump{\omega} = 0$ and $\jump{\partial_y\omega} = 0$. 
The masked vorticities $\omega_1$ and $\omega_2$ each evolve according to
\begin{equation}
    \partial_t \omega_i + (\mathbf u_1 + \mathbf u_2)\cdot \boldsymbol \nabla \omega_i = \frac{1}{Re}\Delta \omega_i - n \cos (ny),
    \label{eqn:vort_coupled}
\end{equation}
on their respective domains. 
Here the velocity has been decomposed into components induced by the vorticity on each half domain, with $\mathbf u_i := (-\hat{\mathbf x} \partial_y + \hat{\mathbf y} \partial_x) \Delta^{-1}\omega_i$. 

We then quantify coupling between the vorticity fields by the action of the induced velocity from one field onto another, defining a \emph{coupling coefficient} $j \to i$:
\begin{equation}
    c_{ij} := \| \mathbf u_j(\omega_j) \cdot \boldsymbol \nabla \omega_i\|,
    \label{eqn:coupling_defn}
\end{equation}
where $\|\bullet\| := [(8\pi^2)^{-1}\iint |\bullet|^2 d^2\mathbf x]^{1/2}$ and the definition (\ref{eqn:coupling_defn}) allows for examination of `self interaction' when $i = j$.

It is helpful to consider the normalized version of (\ref{eqn:coupling_defn}): $C_{ij}:= c_{ij}/(c_{11}+c_{22}+c_{12}+c_{21})$ with $C_{11}+C_{22}+C_{12}+C_{21}=1$ by construction. 

The cross-coupling terms, $C_{12}$ and $C_{21}$, were computed for snapshots on a long turbulent trajectory (with discrete shifts applied to $\omega$ in $y$ such that the $C_{12} + C_{21}$ was minimized) and a joint probability density function (pdf) of these variables is reported in figure \ref{fig:coupling}(a).
These results imply that the turbulence does indeed spend most of its time in a `weakly coupled' state due to the decay of the induced velocity away from the source regions.
Examples of particularly `weak', `strong' and `one-way' coupled fields are visualized in panels (b)-(d) of figure \ref{fig:coupling}, where we report contours of $\mathbf u_j(\omega_j) \cdot \boldsymbol \nabla \omega_i$, along with the driving vorticity fields. 
The weakest-coupled example (panel (b)) is associated with particularly intense vorticity fields with many closely-packed vortices providing a shielding effect for the other half of the domain.


\begin{figure}
\includegraphics[width=1\columnwidth]{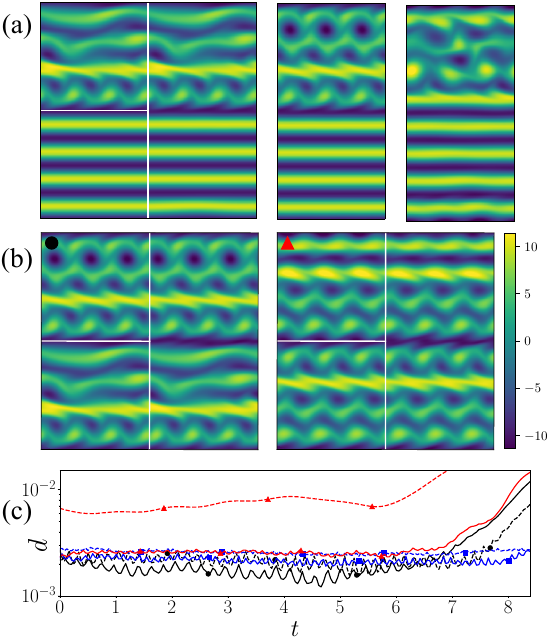}
\caption{
Exact coherent states in the tall $2\pi \times 4\pi$ box.
(a) Snapshots of out-of-plane vorticity for RPOs formed from a combination of a small box ($2\pi\times 2\pi$) RPO with the laminar solution. The first panel includes the `starting' pair of solutions alongside the converged state. 
(b) Snapshots of out-of-plane vorticity for trajectories shadowing two-tori in the $2\pi \times 4\pi$ box. Left panels show snapshots from the pair of RPOs used to initialize the optimization, right panel is a snapshot from the final state. 
(c) Masked near-recurrence for the solutions in (b). 
Colors match the symbols on the panels in (b); dashed lines are top solution, solid lines the bottom. Symbols in (c) identify integer multiples of the fundamental period within the masked subdomains (blue solution shown in SI). 
}
\label{fig:big_fig1}
\end{figure}



Large-box structures are sought via gradient-based minimization of objective functionals of the form
\begin{equation}
    \mathscr L(\mathbf u, T, \alpha) = \frac{1}{2}\|\mathscr M( \mathscr T^{\alpha} \mathbf f^T(\mathbf u) - \mathbf u)\|^2 + \text{constraints}.
    \label{eqn:base_loss}
\end{equation}
Here $\mathscr M$ is a smooth function which windows the velocity field in the vertical, $\mathscr T^{\alpha}:\mathbf u(x, y) \to \mathbf u(x + \alpha,y)$ is a continuous shift operator and $\mathbf f^{T}$ is the flow map with $T$ the target time. 
The forward fields satisfy the Navier-Stokes equations which is enforced in (\ref{eqn:base_loss}) via constraints
$
\int_0^T \left\langle{\bf v}^\dag \cdot \left( \partial_t{\bf u}+({\bf u}\cdot\boldsymbol \nabla){\bf u}+\boldsymbol \nabla p- Re^{-1}\Delta {\bf u} - \hat{\mathbf x}\sin (ny)\right)\right\rangle\,dt
$ 
and 
$
\int_0^T\left\langle\pi^\dag\boldsymbol\nabla\cdot{\bf u}\right\rangle\, dt\,
$ 
where $\langle \bullet \rangle := (8\pi^2)^{-1} \iint \bullet \, d^2\mathbf x$ and $\mathbf v^{\dagger}$, $\pi^{\dagger}$ are adjoint velocity and pressure fields respectively.
We consider various versions of (\ref{eqn:base_loss}) in this paper, changing both the masking profile $\mathscr M$ and adding new terms to seek near recurrence on local patches of the domain.

We solve the optimization problems described below using variational techniques which relate the gradients of the objective functional to the adjoint field $\mathbf v^{\dagger}(t=0)$ (see the Supplemental Information). 
The optimization is complemented with classical Newton-Krylov techniques for convergence of exact RPOs and also as a minimization approach for non-$T$-periodic states (see SI). 
Forward and backward time integration is performed with the Dedalus solver \cite{dedalus} with a Fourier decomposition in both $x$ and $y$ ($N_x \times N_y = 128\times256$) while time integration is performed with a 3rd-order operator-splitting Runge-Kutta scheme. 
Gradients are passed to an AdaGrad optimizer \cite{duchi2011} (see SI for details of specific hyperparameters).


We first attempt to tile RPOs from the extensive library assembled in \cite{Cleary2025} with the laminar solution to form new RPOs. 
To do this, the masking function $\mathscr M$ is set to the identity function. We also found it necessary to modify equation (\ref{eqn:base_loss}) to a multi-point shooting form which splits the full loop $t\in[0,T]$ into subintervals (see SI). 
Following optimization, candidate RPOs were passed to a Newton solver and converged to a relative error $<10^{-10}$.


Three examples of the 25 successful RPO-laminar convergences are reported in figure \ref{fig:big_fig1}(a). 
For all converged states, the candidate small-box RPOs tend to be very high-dissipation, with many consisting of counter-propagating rows or `jets' of vorticity which shield the laminar patches from distortion (though note the weak distortion of the laminar stripes in the third panel. 
The converged RPO-laminar tiles are also shown in green in figure \ref{fig:coupling}, with the majority sitting away from the turbulent attractor in an extremely weakly coupled regime.


We next attempt to combine pairs of (non-laminar) RPOs to form two-tori, with the expectation being that weak coupling allows for structures which are approximately periodic on half of the domain. 
The loss function (\ref{eqn:base_loss}) is therefore adjusted to seek near recurrence in two distinct regions:
\begin{align}
    \mathscr L(\mathbf u, T_1, &\alpha_1, T_2, \alpha_2) = \frac{1}{2}\|\mathscr M_1( \mathscr T^{\alpha_1} \mathbf f^{T_1}(\mathbf u) - \mathbf u)\|^2 \nonumber \\
    + \; &\frac{1}{2}\|\mathscr M_2( \mathscr T^{\alpha_2} \mathbf f^{T_2}(\mathbf u) - \mathbf u)\|^2 + \; \text{constraints}.
    \label{eqn:torus_loss}
\end{align}
where the mask $\mathscr M(\mathbf u) \equiv \mathbb L(M(y) \mathbf u)$, where $M(y)$ is the vertical windowing function and $\mathbb L$ is a Leray projection.
The (differentiable) window function $M(y)$ is formed by a double vertical convolution of a pair of square waves (see SI).
We incrementally increase the target time over which (\ref{eqn:torus_loss}) is minimized: 
once $\mathscr L$ is reduced such that the relative error $\varepsilon_i:=\|\mathscr M_i( \mathscr T^{\alpha_i} \mathbf f^{T_i}(\mathbf u) - \mathbf u)\|/\|\mathscr M_i \mathbf u\| \leq 0.01$ in each region, we then restart the process with $T_i \to 2 T_i$. 
All results presented here were converged below this tolerance to at least two fundamental periods in both halves of the domain.



\FloatBarrier
\begin{figure}[t] 
  \centering
\includegraphics[width=0.8\columnwidth]{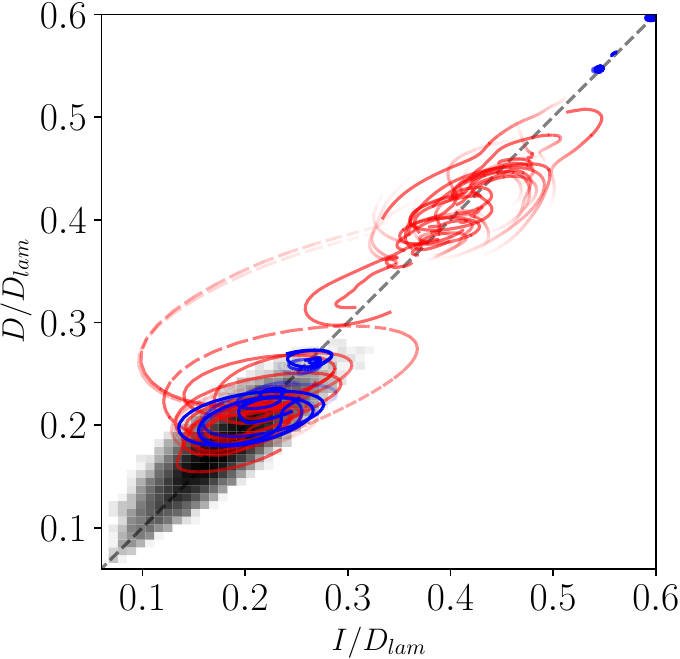}%
\vspace{4pt}
\caption{
Production/dissipation ($I$ and $D$ respectively) in the $2\pi \times 4\pi$ domain.
Gray histogram is a pdf of a long ($4000$ advective time units) turbulent orbit,  which runs logarithmically from 0.4 to 275.8. 
Also shown are 
the candidate two-tori (blue) and some example `turbulent' trajectories which shadow an RPO in a masked region of the domain (red). The dashed/faded components of these curves span the first half of the optimization period $nT$ for tiles with turbulence and $\min(n_1T_1,n_2T_2)$ for RPO-RPO tiles.
}
\label{fig:PD}
\end{figure}
\begin{figure}[t!] 
    \centering
   \includegraphics[width=\columnwidth]{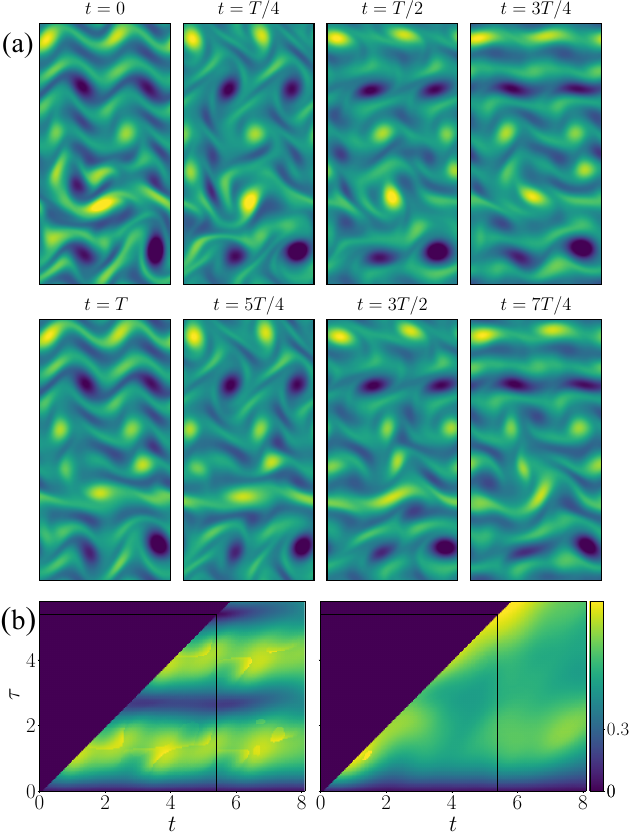}
\caption{
RPO shadowing in a subdomain. (a) Out-of-plane vorticity above two periods of the RPO used to construct the solution (contours run $-11.4 \leq \omega \leq 11.4$). (b) Masked autorecurrence $R(t,\tau)$ in the top (left panel) and bottom (right panel) halves of the domain.}
\label{fig:turb_tile} 

\end{figure}

We apply  this procedure for all combinations of higher-dissipation RPOs that successfully tiled with the laminar state.
From this search we obtain six trajectories which converge to a relative error within the masked regions of less than 0.01 for both regions over at least two periods. 
Some examples are included in figure \ref{fig:big_fig1}(b), with figure \ref{fig:big_fig1}(c) showing the corresponding evolution of the minimum relative distance from the original $2\pi\times2\pi$ states vertically stacked on the $4\pi$ domain,
\begin{equation}
d_i(t) = \min_{\alpha,t'} \norm{\mathscr{M}_i\left(\ub_{4\pi}(t)-\mathscr{T}^\alpha\ub_{2\pi,i}(t')\right)}/\norm{\ub_{2\pi,i}(t')}\,, 
\end{equation}
where $\ub_{4\pi}$ and $\ub_{2\pi,i}$ are the tiled $2\pi\times4\pi$ solution and the $2\pi\times2\pi$ RPOs ($i \in \{1, 2\}$) used to generate it in the corresponding half domain, respectively, 
over several periods. 
The distinct fundamental frequencies, $2\pi / T_1$ and $2\pi /T_2$, and low values of $d_i$ extending over several Lyapunov times ($t_{\lambda} \sim 2$ here) are strongly suggestive of shadowing of underlying two-tori.

Similar to the laminar tiles discussed above, the solutions reported in figure \ref{fig:big_fig1} are all extremely weakly coupled, and with vortices contained in distinct vertical bands. 
However, five dynamically relevant candidate tori were also obtained with coupling coefficients and dissipation ranges which are similar to values obtained on a turbulent orbit.
These solutions are overlayed on the turbulent attractor visualized in figure \ref{fig:coupling}. 
In contrast to the states reported in figure \ref{fig:big_fig1}, which are difficult to distinguish from the starting small-box RPOs, these dynamically relevant tori are altered in the optimization process. Example visualizations are included in the SI, where we also report full time evolutions to confirm the masked autorecurrence in each half of the spatial domain.

The various RPOs/tori described above are all visualized in a production-dissipation diagram in figure \ref{fig:PD} (the production $I := \langle u \cdot \sin(ny)\rangle$). 
The `dynamically relevant' two tori which combine distinct high-dissipation RPOs overlap the turbulent attractor (see blue curves in the lower panel). 
In addition, we also visualize in red a number of turbulent trajectories which \emph{locally} shadow a small box RPO. 
These events were found via mimization of equation (\ref{eqn:base_loss}) with the mask applied to a single region with the dynamics outside essentially unconstrained.



This approach is very effective and produces $d \leq 0.01$ over $>2 $ full periods of the underlying RPO for many RPOs in our library, producing 32 (16 unique RPOs) such solutions from the 21 RPOs used to construct two-tori. An example trajectory is reported in figure \ref{fig:turb_tile}(a), where near recurrence in the top half of the domain can be clearly seen (compare snapshots vertically at equivalent times modulo the period $T$). 
In panel (b) we also show the masked autorecurrence function, 
$
    R(t,\tau) = \min_{\alpha}\|\mathscr M(\mathscr{T}^{\alpha}({\bf u}(t)-{\bf u}(t-\tau))\| / \|\mathscr M{\bf u}\|,
$
which confirms the shadowing in the top half while the lower half of the domain shows no temporal autocorrelation beyond $\tau \approx 1$.


In this Letter we have shown that RPOs converged in a `small-box' calculation can be combined to form new solutions of the Navier-Stokes equation on a larger spatial domain. 
This spatial tiling has produced exact RPOs, two-tori involving weakly interacting substructures and more general `turbulent' orbits which shadow exact solutions on patches of the domain. 
These latter structures are in line with the view of turbulence as a `spatiotemporal tapestry' of connected states.
The algorithms employed here are general and easily adapted to three-dimensional problems, where a long-standing challenge has been the identification of RPOs with distinct substructures. 


\bibliography{refs.bib}

\end{document}


\title{Exact coherent structures as building blocks of turbulence on large domains \\ Supplemental information}

\author{Dmitriy Zhigunov}
\email{dmitriy.zhigunov@ed.ac.uk}
\author{Jacob Page}%
 \email{jacob.page@ed.ac.uk}
\affiliation{%
 School of Mathematics and Maxwell Institute for Mathematical Sciences, University of Edinburgh,
Edinburgh, EH9 3FD, UK
}%
\maketitle

\section{Further details on methodology}
\subsection{Multi-point gradient descent}
\label{sec:multi_pt}
All Navier-Stokes solutions reported in the main manuscript are found through minimization of a scalar loss function (for relative periodic orbits --RPOs-- this is an initial step performed prior to application of an exact root-finding algorithm). 
Convergence is aided in some cases by extending the `single point' loss function of \cite{Page2024} to include $N \geq 1$ points on the trajectory:
\begin{align}
    \mathscr L[\{\mathbf u_j\}_{j=0}^{N-1}, \Delta T, \alpha] = 
    &\frac{1}{2}\sum_{j=0}^{N-1}\|\mathscr M({\mathbf  f}^{\Delta T}(\mathscr{T}^{\alpha \delta_{j(N-1)}}{\mathbf u}_{j})-{\mathbf u}_{(j+1)\bmod N})\|^2 \nonumber \\
    - & \sum_{j=0}^{N-1}\int_{0}^{\Delta T}
    \langle \mathbf v^{\dagger}_j \cdot \left(\partial_t \mathbf u + \mathbf u \cdot \boldsymbol \nabla \mathbf u + \boldsymbol \nabla p - Re^{-1}\Delta \mathbf u - \hat{\mathbf x} \sin n y \right)\rangle
     dt \label{eqn:multiloss} \\
     - & \sum_{j=0}^{N-1}\int_{0}^{\Delta T}
    \langle \pi^{\dagger}_j \boldsymbol \nabla \cdot \mathbf u\rangle \, dt 
\end{align}
Note the treatment of the end-point: the $x$-shift is applied only at this point and the final field is compared to the initial condition $\mathbf u_0$.
The notation in (\ref{eqn:multiloss}) follows that in the main paper, with $\mathscr M$ a windowing function and $\mathbf f^t$ the time-forward map associated with the Navier-Stokes equations.
The variables $\mathbf u_j$ are the \emph{initial conditions} on each of the sections of the full orbit, while the period of the underlying solution is $T = N\Delta T$.
Constraints are enforced by $N$ adjoint and pressure fields, $\{\mathbf v_j(\mathbf x, t)\}$ and $\{\pi_j(\mathbf x, t)\}$, which are functions of space and time. 
To aid the discussion below we will mix notation and use $\mathbf u_j(t)\equiv \mathbf f^t (\mathbf u_j)$.
Note that in most cases we employ a single-point loss with $N=1$. Multi-point shooting was necessary to find the RPOs which result from tiling a small-box RPO with the laminar state.

Gradients with respect to initial conditions are found using the standard definition for the first variation
\begin{align*}
    \delta \mathscr L[\mathbf u, \delta \mathbf u] &:= \lim_{\varepsilon \to 0}\left(\frac{\mathscr L[\mathbf u + \varepsilon\delta \mathbf u] - \mathscr L[\mathbf u]}{\varepsilon}\right) \\
    &:= \int_0^T \bigg \langle \frac{\delta \mathscr L}{\delta \mathbf u} \cdot \delta \mathbf u \bigg \rangle dt.
\end{align*}
Consider first the derivative of the objective functional without constraints with respect to initial condition $i$ in the set $\{\mathbf u_j\}_{j=1}^N$:
\begin{align*}
    &\bigg \langle \frac{\delta}{\delta \mathbf u_i} \frac{1}{2}\sum_{j=0}^{N-1}\| \mathscr M\left(\mathbf f^{\Delta T}(\mathscr T^{\alpha \delta_{j(N-1)}} \mathbf u_j) - \mathbf u_{(j+1) \bmod N} \right)\|^2 \cdot \delta \mathbf u_i \bigg \rangle = \\
    &\bigg \langle \mathscr M(\mathbf f^{\Delta T}(\mathbf u_i) - \mathscr T^{-\alpha \delta_{i (N-1)}} \mathbf u_{i+1}) \cdot \delta \mathbf u_i(T)  \bigg \rangle 
    - \bigg \langle \mathscr M(\mathbf f^{\Delta T}(\mathscr T^{\alpha \delta_{(i-1) (N-1)}}  \mathbf u_{(i-1)\bmod N}) - \mathbf u_{i}) \cdot \delta \mathbf u_i(0)  \bigg \rangle
\end{align*}
where $\delta \mathbf u_i(\Delta T) \equiv \langle \delta_{\mathbf u} \mathbf f^{\Delta T}(\mathbf u_i) \cdot \delta \mathbf u_i(0)\rangle$.

Taking variations of the constraint terms is a simple exercise in integration by parts following the `standard' procedure (e.g. see \cite{kerswell2018}). 
Call $\mathscr L_C$ the `constraint' component of the objective functional (\ref{eqn:multiloss}), then
\begin{align*}
    \int_0^{\Delta T}\bigg\langle  \frac{\delta \mathscr L_C}{\delta \mathbf u} \cdot \delta \mathbf u  \bigg \rangle dt &+ 
    \int_0^{\Delta T}\bigg\langle  \frac{\delta \mathscr L_C}{\delta p} \cdot \delta p  \bigg \rangle dt = \\
    &\hphantom{=}- \int_0^{\Delta T}\bigg\langle \mathbf v^{\dagger}\cdot\left(\partial_t \delta \mathbf u + \mathbf u\cdot \boldsymbol{\nabla} \delta \mathbf u + \delta \mathbf u \cdot \boldsymbol{\nabla} \mathbf u + \boldsymbol{\nabla}\delta p - \frac{1}{Re}\Delta \delta \mathbf u \right) \bigg \rangle dt \\
    &\hphantom{=}- \int_0^{\Delta T}\bigg\langle \pi^{\dagger}\boldsymbol{\nabla}\cdot \mathbf \delta \mathbf u \bigg \rangle dt \\
    &= -\int_0^{\Delta T}\partial_t\langle \mathbf v^{\dagger}\cdot \delta \mathbf u \rangle dt \\
    &\hphantom{=}-\int_0^{\Delta T} \bigg \langle \delta \mathbf u \cdot \left(
    -\partial_t\mathbf v^{\dagger} - \mathbf u\cdot \boldsymbol{\nabla}\mathbf v^{\dagger} + \mathbf v^{\dagger}\cdot (\boldsymbol{\nabla}\mathbf u)^T - \boldsymbol{\nabla}\pi^{\dagger} - \frac{1}{Re}\Delta \mathbf v^{\dagger}
    \right) \bigg \rangle dt \\
    &\hphantom{=}+ \int_0^{\Delta T}\langle \delta p \boldsymbol{\nabla}\cdot \mathbf v^{\dagger}\rangle dt,
\end{align*}
where we show just one of the $N$ contributions to the sum in (\ref{eqn:multiloss}) for illustration and periodic boundary conditions were applied to obtain the second expression.
Integrating the first term on the right hand side (temporal boundary term) and combining with the functional derivatives of the unconstrained objective functional above we obtain 
\begin{align}
    \frac{\delta \mathscr L}{\delta \mathbf u_j(\Delta T)} &= \mathscr M(\mathbf f^{\Delta T}(\mathbf u_j) - \mathscr T^{-\alpha\delta_{j(N-1)}}\mathbf u_{(j+1)\bmod N}) - \mathbf v^{\dagger}_j(\Delta T), \label{eqn:opt_final}
    \\
    \frac{\delta \mathscr L}{\delta \mathbf u_j(0)} &= \mathbf v^{\dagger}_j(0) - \mathscr M(\mathbf f^{\Delta T}(\mathscr T^{\alpha \delta_{(j-1) (N-1)}}  \mathbf u_{(j-1)\bmod N}) - \mathbf u_{j}). \label{eqn:opt_initial}
\end{align}
The more straightforward derivatives are 
\begin{align}
   \frac{\partial \mathscr L}{\partial \Delta T} &= \sum_{j=0}^{N-1}\bigg \langle \mathscr M(\mathbf f^{\Delta T}(\mathscr T^{\alpha \delta_{j(N-1)}} \mathbf u_j) - \mathbf u_{(j+1)\bmod N}) \cdot \frac{\partial \mathscr M \mathbf f^{\Delta T}(\mathscr T^{\alpha  \delta_{j(N-1)}} \mathbf u_j)}{\partial \Delta T} \bigg \rangle,  \label{eqn:opt_T}
   \\
   \frac{\partial \mathscr L}{\partial \alpha} &=
   \bigg \langle \mathscr M(\mathbf f^{\Delta T}(\mathscr T^{\alpha} \mathbf u_{N-1}) - \mathbf u_{0}) \cdot \frac{\partial \mathscr M \mathbf f^{\Delta T}(\mathscr T^{\alpha} \mathbf u_{N-1})}{\partial \alpha} \bigg \rangle.
   \label{eqn:opt_shift}
\end{align}

Setting the first variation to zero then naturally leads to the classical `adjoint looping' algorithm (see for instance \cite{Schmid2007,kerswell2018}). 
In this approach, given an initial guess $(\{\mathbf u_j^0\}_{j=0}^{N-1}, \Delta T^0, \alpha^0)$, each initial condition is marched forward $\Delta T^0$, and the final condition on the adjoint fields are obtained from setting the variation (\ref{eqn:opt_final}) to zero:
\begin{equation*}
    \mathbf v^{\dagger0}_j(\Delta T^0) = \mathscr M(\mathbf f^{\Delta T^0}(\mathbf u_j^0) - \mathscr T^{-{\alpha^0}\delta_{j(N-1)}}\mathbf u^0_{(j+1)\bmod N}).
\end{equation*}
The adjoint equations, 
\begin{subequations}
    \begin{equation}
        \partial_t\mathbf v^{\dagger} + \mathbf u\cdot \boldsymbol{\nabla}\mathbf v^{\dagger} - \mathbf v^{\dagger}\cdot (\boldsymbol{\nabla}\mathbf u)^T = - \boldsymbol{\nabla}\pi^{\dagger} - \frac{1}{Re}\Delta \mathbf v^{\dagger},
    \end{equation}
    \begin{equation}
        \boldsymbol \nabla \cdot \mathbf v^{\dagger} = 0,
    \end{equation}
    \label{eqn:adjoint}
\end{subequations}
are then marched backwards in time to $t=0$ which allows for computation of all derivatives with respect to the guess (equations \ref{eqn:opt_initial}, \ref{eqn:opt_T}, \ref{eqn:opt_shift}).
We pass these gradients to the AdaGrad optimizer \cite{duchi2011} to update the guess and iterate for a finite number of steps or until some convergence criteria is reached (detailed below). 
In most cases we then apply a Newton-Hookstep minimization which we also describe in \S\ref{sec:hookstep} below.

In the AdaGrad optimization we employ a different `learning rate' $\eta$ (an initial step size for gradient descent) depending on the specific problem: 
For the exact RPOs which are tiled with the laminar solution we use $\eta=0.025$, a value of $\eta = 0.001$ was used for the two-tori and $\eta = 0.002$ was used to identify turbulent trajectories which locally shadow a small-box RPO.

\subsection{Spatial masking functions}
\label{sec:spatial_mask}
\begin{figure}
\includegraphics[width=\textwidth]{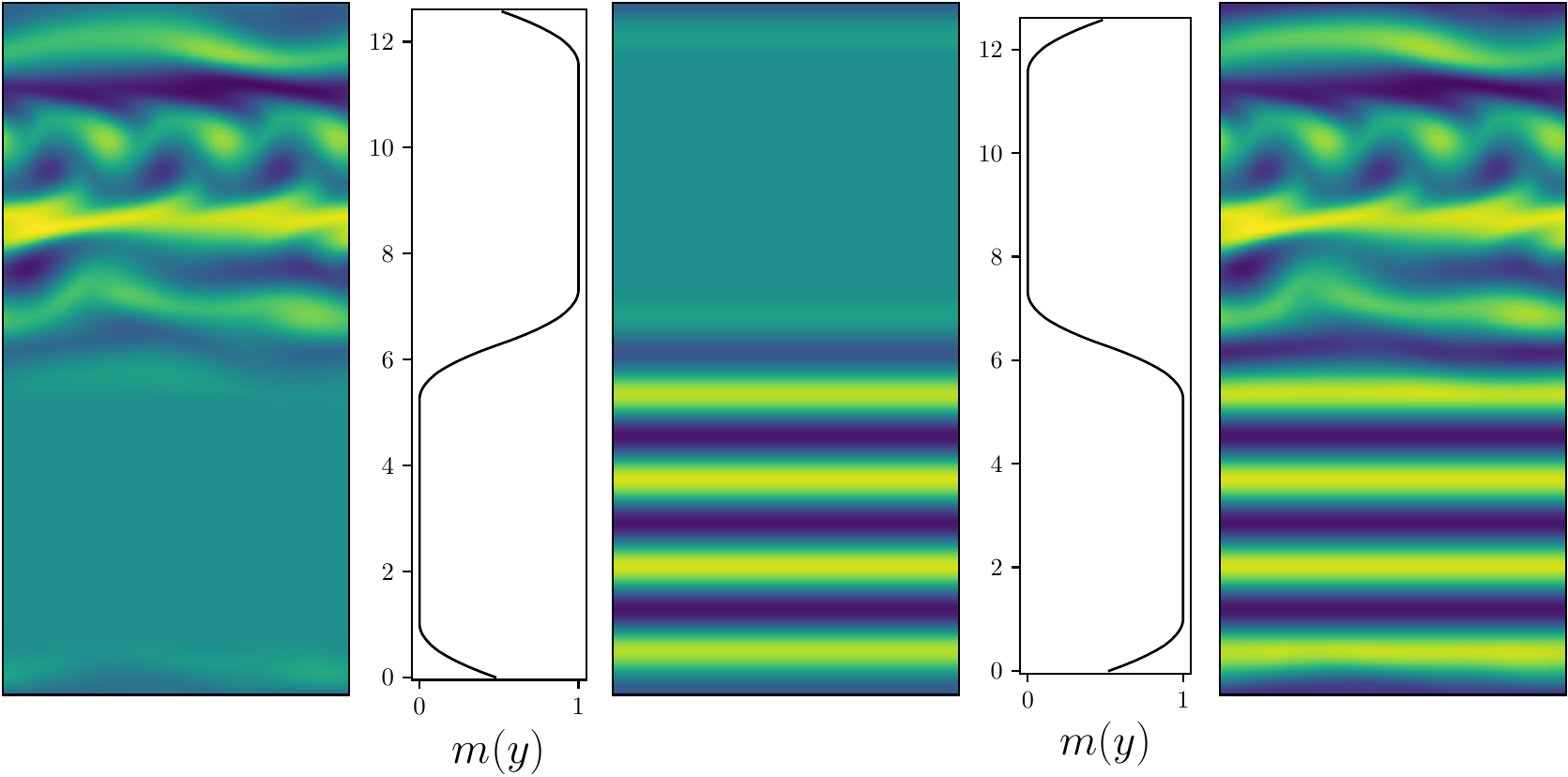}
\caption{\label{fig:tiling_schem} A schematic of how initial guesses for tiled solutions are generated, showing the two smoothed fields from initial $2\pi$ ECSs which are then summed together to generate an intial guess for the optimiser. }
\end{figure}

As described in the main text, the mask $\mathscr M(\mathbf u)$ used to measure spatially-localized near recurrence is the combination of (i) multiplication by a windowing function $M(y)$ and (ii) application of a Leray projection $\mathbb L$ to ensure that the solution remains solenoidal, so that $\mathscr M(\mathbf u) \equiv \mathbb L(M(y) \mathbf u)$. 
The mask is also used to generate initial guesses from the $2\pi \times 2\pi$ solutions which seed the algorithm on the $2\pi \times 4\pi$ domain.

The windowing function  used in this work is the convolution of several `rectangular' functions 
\begin{equation}
    \Pi(y/Y) := \begin{cases} 1 \quad \text{if} \quad y\in[0,Y], \\
    0 \quad \text{otherwise}.
    \end{cases}
\end{equation}
For illustration, consider the mask which extracts a solution centred about $y=\pi$, 
$
    M_0(y) := \Pi(y/ 2\pi)
$.
To form the differentiable mask $M(y)$ used in $\mathscr M$, we convolve $M_0(y)$ twice with a second, narrow rectangular function, $\Pi(y/a)$, where $a = 5\pi /16$.
Examples of this function -- suitably translated in $y$ -- are provided in figure \ref{fig:tiling_schem}.

Figure \ref{fig:tiling_schem} also illustrates the procedure by which initial fields on the $2\pi\times 4\pi$ domain are generated from exact $2\pi \times 2\pi$ solutions.
First, each small-box solution is copied twice in the vertical, where they remain exact solutions in the new $2\pi \times 4\pi$ domain. 
The masks $\mathscr M_j(\mathbf u_j)$ are then applied to each solution, which includes a Leray projection, centred either on $y= \pi$ or $y=3\pi$ and resulting divergence-free fields are summed to produce the initial guess for the large-box solution.

\subsection{Optimization with Newton-Krylov methods}
\label{sec:hookstep}
Following the application of the adjoint-optimization procedure described above, we apply a variant of the Newton-Krylov-Hookstep algorithm \cite{viswanath2007,chandler2013} to either refine the structure (in the case of two-tori or local shadowing of RPOs by turbulence) or converge it exactly (for small-box-RPO and laminar combinations).
Here we give a high-level summary of the approach before describing the invariant solutions, where the algorithm differs in the design of the specific Krylov subspaces.

Our loss functions are typically of the form
\begin{equation}
    \mathscr L = \frac{1}{2}\| \mathscr M(\mathbf f^{T}(\mathscr T^{\alpha}\mathbf u) - \mathbf u)\|^2 + \text{constraints},
\end{equation}
which is the norm of (masked) function whose roots we seek when converging exact RPOs, i.e. $\mathscr M(\mathbf F(\mathbf u, \alpha, T))$, where
\begin{equation}
    \mathbf F(\mathbf u,\alpha, T) = \mathbf f^{T}(\mathscr T^{\alpha}\mathbf u) - \mathbf u.
    \label{eqn:root_fn}
\end{equation}
We can consider an update to an initial guess, $(\mathbf u, T, \alpha) \to (\mathbf u + \delta \mathbf u, T + \delta T, \alpha + \delta \alpha)$. 
Inserting this into (\ref{eqn:root_fn}) and truncating at first order yields the classic Newton-Raphson algorithm, 
\begin{equation}
    \mathscr M \mathbf J \delta \mathbf w = -\mathscr M \mathbf F(\mathbf w) 
\end{equation}
where we have introduced $\mathbf w := (\mathbf u, T, \alpha)$ for convenience and 
\begin{equation*}
    \mathbf J := \begin{bmatrix}
        \boldsymbol \nabla_{\mathbf u} \mathbf f^{T}(\mathscr T^{\alpha} \mathbf u) - \mathbf I & \partial_T\mathbf f^{T}(\mathscr T^{\alpha} \mathbf u) & \partial_{\alpha}\mathbf f^{T}(\mathscr T^{\alpha} \mathbf u)
    \end{bmatrix},
\end{equation*}
is the usual definition of the Jacobian. Without the mask, we need to introduce additional constraints associated with temporal/spatial drift to make the problem well-posed. 
With the mask included, the problem is underdetermined (if the mask exactly zeros the solution in parts of the spatial domain). 
However, we can still solve for an update direction in a Krylov subspace $\delta \hat{\mathbf w}_n \in \mathscr K^n$ by solving the minimization (GMRES) problem:
\begin{equation}
    \delta \hat{\mathbf w}_n := \argmin_{\delta \mathbf w_n\in \mathscr K^n}\|\mathscr M(\mathbf J\delta \mathbf w_n + \mathbf F(\mathbf w))\| \quad \text{s.t.} \quad \|\delta \hat{\mathbf w}_n\| \leq \lambda.
\end{equation}
The subspace $\mathscr K^n$ is built from the iterative action of $\mathbf J$ on a (set of) starting vector(s) which varies between the different invariant solutions (see below). 
The size of the subspace is increased incrementally until $\|\mathscr M(\mathbf J\delta \hat{\mathbf w}_n + \mathbf F(\mathbf w))\|/\|\mathbf w\|$ falls below a specified tolerance, usually $0.01$, or until a maximum of steps $n_G$ is reached. If tolerance is not achieved for a given Newton step, $n_G \to n_G + 7$ at the next iteration. At initialization $n_G = O(10)$.
The inclusion of the Hookstep constraint connects the method to the optimization: the size of the trust region $\lambda$ is reduced until $\|\mathscr M \mathbf F(\mathbf w + \delta \hat{\mathbf w}_n)\| \leq \|\mathscr M \mathbf F(\mathbf w)\|$, which ensures that $\mathscr L$ decreases. 
\subsection{Convergence details and Krylov subspaces}
\subsubsection{RPO-laminar tiles}
Initial guesses for $2\pi \times 4\pi$ RPOs were constructed by patching together a small-box RPO with the laminar solution, as described in \S\ref{sec:spatial_mask}.
The initial guess was marched forward by the period of the small box solution, $T_{2\pi}$, and multi-point gradient descent was then applied if the initial loss $\mathscr L \leq 0.6$.
Multi-point optimization with AdaGrad was performed with $N = \text{floor}(T+1)$ points $\{\mathbf u_j\}$ for 200 steps.
Because we anticipate convergence of an RPO on the large domain, the mask in the optimization (see equation \ref{eqn:multiloss}) was set as the identity function $\mathscr M \equiv \mathscr I$. 
A multi-point Newton method was then applied until a relative error $\varepsilon \leq 10^{-4}$, before a final single-point Newton method was used to reduce $\varepsilon \leq 10^{-10}$ due to the increasingly large Krylov subspaces required in the multipoint version. 
The `standard' subspace of $\mathscr K^n = \text{span}\{\mathbf F(\mathbf w), \mathbf J \mathbf F(\mathbf w), \cdots, \mathbf J^{n-1}\mathbf F(\mathbf w)\}$ was used here.


\subsubsection{Two-tori from pairs of RPOs}
Guesses for two-tori are formed from tiled combinations of small-box RPOs. 
The search space for solutions was initially limited to the 25 small-box solutions which successfully tiled with the laminar state, before being expanded to a wider set containing an additional 25 RPOs which all had periods $T<3$ and mean dissipation rates $\overline{D}/D_{l}<0.7$.

For optimization, we sought to minimize the loss
\begin{align}
    \mathscr L(\mathbf u, T_1, \alpha_1, T_2, \alpha_2) = &\frac{1}{2}\|\mathscr M_1( \mathscr T^{\alpha_1} \mathbf f^{T_1}(\mathbf u) - \mathbf u)\|^2 
    + \; \frac{1}{2}\|\mathscr M_2( \mathscr T^{\alpha_2} \mathbf f^{T_2}(\mathbf u) - \mathbf u)\|^2 
    + \; \text{constraints}
    \label{eqn:torus_loss}
\end{align}
(note a single point, $\mathbf u$, rather than the multi-point method above) using an AdaGrad optimization for $2\times 100$ iterations (we restarted after the first 100). 
We then applied the Hookstep optimization described in \S\ref{sec:hookstep} on the function
\begin{equation}
    \mathbf T(\mathbf u, T_1, \alpha_1, T_2, \alpha_2) := \mathscr M_1 \mathbf F(\mathbf u, T_1, \alpha_1) + \mathscr M_2\mathbf F(\mathbf u, T_2, \alpha_2),
    \label{eqn:torus_root}
\end{equation}
where $\mathbf F$ is defined above (equation \ref{eqn:root_fn}). 
The choice of Krylov subspace in which the GMRES is performed is key:
\begin{align}
    \mathscr K^{2n + 5} = \text{span}
    \{
    &\mathbf T, 
    \hat{\mathbf e}_{T_1}, 
    \hat{\mathbf e}_{\alpha_1}, 
    \hat{\mathbf e}_{T_2}, 
    \hat{\mathbf e}_{\alpha_2}, 
    \mathbf F(\mathbf u, T_1, \alpha_1),
    \mathbf F(\mathbf u, T_2, \alpha_2), \nonumber \\
    &\mathbf J_1 \mathbf F(\mathbf u, T_1, \alpha_1), 
    \mathbf J_2 \mathbf F(\mathbf u, T_2, \alpha_2),
    \dots ,
    \mathbf J_1^{n-1} \mathbf F(\mathbf u, T_1, \alpha_1), 
    \mathbf J_2^{n-1} \mathbf F(\mathbf u, T_2, \alpha_2)
    \}
    \label{eqn:torus_subspace}
\end{align}
where $\mathbf J_i := \boldsymbol \nabla_{\mathbf u, T_i, \alpha_i}\mathbf F(\mathbf u, T_i, \alpha_i)$.
The Jacobian-vector products which span the subspace are evaluated using a simple finite-difference approach (see \cite{chandler2013}).
While the Hookstep minimization (\S \ref{sec:hookstep}) seeks near recurrence in the masked regions (equation \ref{eqn:torus_root}), the Krylov subspace does not involve masking $\mathbf F$ on either subdomain. 

\begin{figure}
\centering
\includegraphics[width=6.5in]{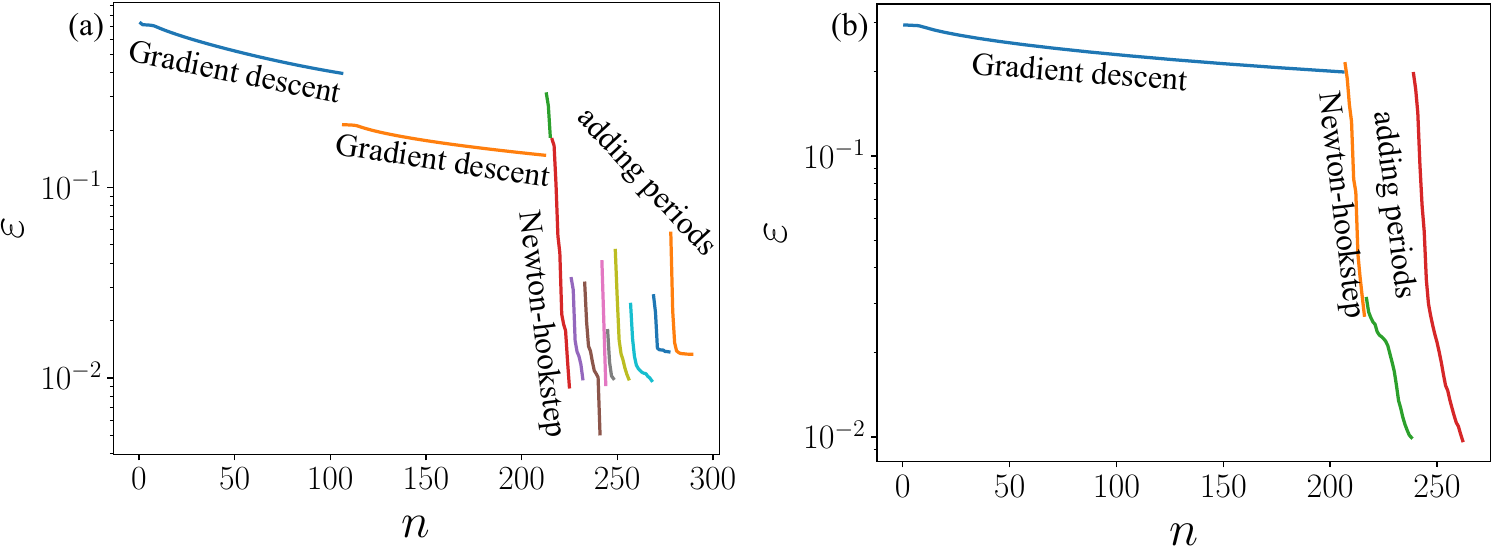}
    \caption{The convergence rates for an example of (a) a two-torus constructed from a pair of small box RPOs 6 (symmetry number 4 and 16 -- see \S\ref{sec:tables} for explanation of this terminology)  and (b) localized shadowing of a small-box RPO by a large box turbulent orbit  corresponding to symmetry number 2 of RPO 6. 
    In both figures, the first Newton convergence is broken up into two segments; a smoothing process is performed between them via time integration for $t=0.1$. In the torus example, the last two added periods do not converge to 1\%, but were included as an example of how the Newton optimisation can stagnate.}
\label{fig:convergence_rates} 
\end{figure}
The minimization is first performed beginning with the fundamental periods of both RPOs, $T_1$ and $T_2$.
Once optimised to a relative error of $\varepsilon := \|\mathbf T\| /\|(\mathscr M_1+\mathscr M_2)\mathbf u\| \leq 10^{-2}$, subsequent periods were added incrementally, and the process repeated. 
The convergence process for an example RPO-RPO torus guess is reported in figure \ref{fig:convergence_rates}.
In this example, we are unable to optimize indefinitely from the initial condition, with the process stalling after $n_T = n_1 + n_2 = 7$ of additions to the periods in the loss, $n_1T_1$ and $n_2T_2$. The final two attempts to add periods stall and are included in the figure for demonstrative purposes.

\subsubsection{Localized shadowing of RPOs by turbulence}
Initial guesses for turbulent trajectories which locally shadow a small-box RPO (`RPO-turbulence tiles') were generated by combining a snapshot from a $2\pi \times 4\pi$ trajectory with a $2\pi \times 2\pi$ RPO using the masking function via the procedure described in \S\ref{sec:spatial_mask}.
We proceeded to optimize if the initial relative (masked) error within the `RPO' region ($2\pi \lesssim y \lesssim 4\pi$) was $\varepsilon := \|\mathscr M(\mathbf f^T(\mathscr T^{\alpha}\mathbf u) - \mathbf u)\|/\|\mathscr M(\mathbf u)\| \leq 0.3$. 


\
The AdaGrad optimization described in \S\ref{sec:multi_pt} was then applied (with a single point, $N=1$) for 200 iterations prior to the Hookstep optimization described in \S\ref{sec:hookstep}. 
The function used in the constrained GMRES was the simple single-masked trajectory, $\mathscr M \mathbf F_0(\mathbf u, T, \alpha) = \mathscr M(\mathbf f^T(\mathscr T^{\alpha} \mathbf u) - \mathbf u)$.

In this problem we perform the Hookstep optimization within a Krylov subspace
\begin{equation}
    \mathscr K^{n+3} = \text{span}\{\mathscr M\mathbf F_0, \hat{\mathbf e}_T, \hat{\mathbf e}_{\alpha}, \mathbf F_0, \mathbf J \mathbf F_0, \dots, \mathbf J^{n-1}\mathbf F_0\},
\end{equation}
where as above $\mathbf J:= \boldsymbol \nabla_{\mathbf u, T, \alpha} F_0$.
Similar to the two-torus approach adopted above, optimization was performed over additional periods once a relative masked error $\varepsilon \leq 10^{-2}$ was obtained. 
An example of the convergence of this approach is reported in figure \ref{fig:convergence_rates}(b).




\begin{figure}
\centering
\includegraphics[width=6.5in]{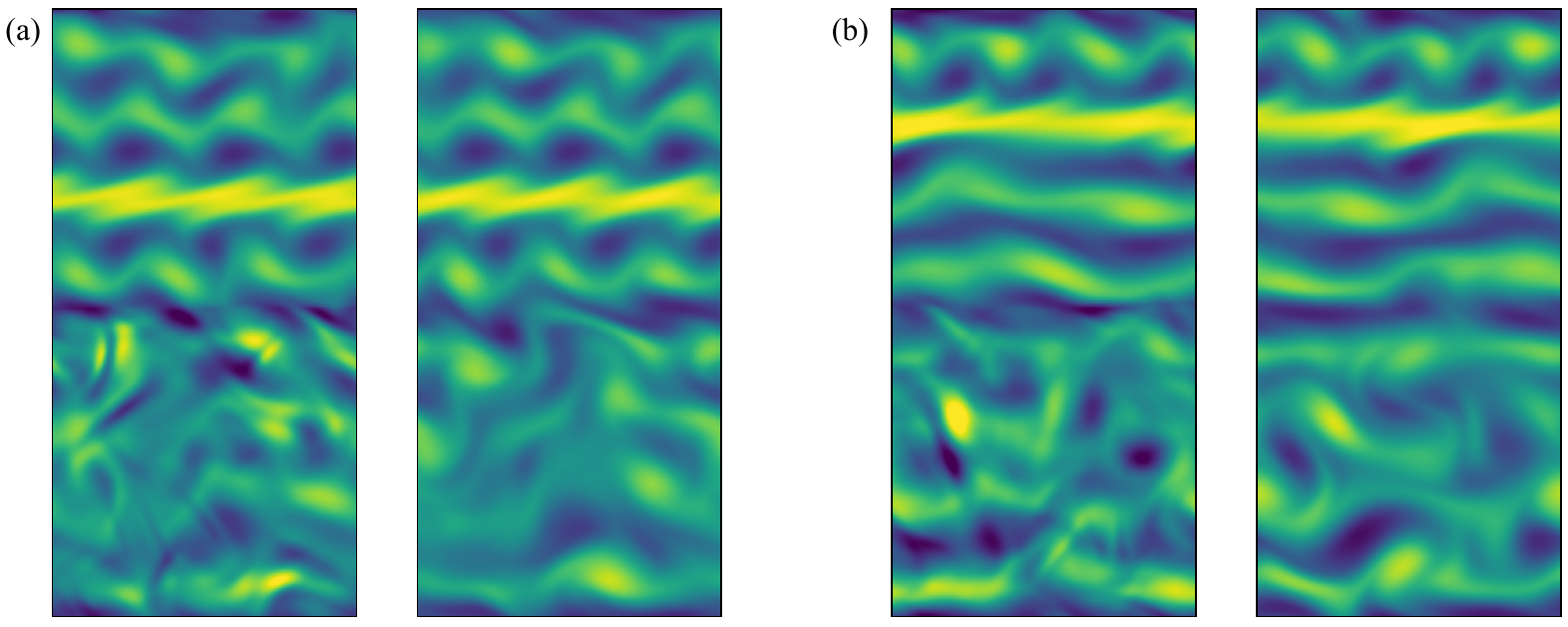}
\caption{
Initial conditions (vorticity contours running $-11.4\leq \omega \leq 11.4$ ) following optimization for large-box trajectories locally shadowing small box RPOs (a) number 6, symmetry number 2 and (b) number 98, symmetry number 0.
The optimization results in some unphysical noise in the initial field (left panels in (a) and (b)) which rapidly dissipates; right panels of (a) and (b) show the field evolved by a half period $T/2$ of the small-box RPO.
}
\label{fig:smoothing} 
\end{figure} 
Notably, this algorithm does not constrain the initial velocity field to be on the attractor, and most initial states converged contain unphysical features. However, these features are quickly smoothed out by time evolution (see example in figure \ref{fig:smoothing}), with the trajectory then locally shadowing the target small-box RPO for multiple periods.

\section{Net vertical flux from adjoint evolution}
\begin{figure}
\includegraphics[width=0.5\textwidth]{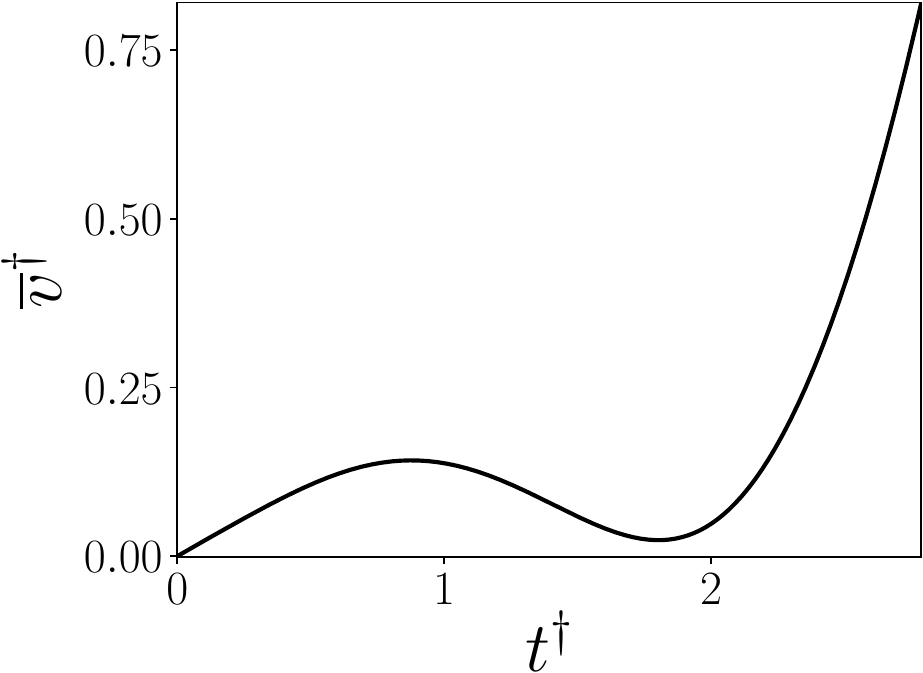}
\caption{\label{fig:mean_v} 
Example evolution of the streamwise-averaged adjoint vertical velocity $\overline{v^\dag}:=(2\pi)^{-1}\int_0^{2\pi} v^{\dagger}(x,y,t) dx$, plotted in terms of the backwards time $t^\dag:=T-t$. The vertical flux is $y$-independent. The corresponding forward velocity field $\ub$ was integrated from a random, smooth initial condition.}
\end{figure}

In the optimization approach presented in \cite{Page2024}, which minimized loss functions of a similar form to (\ref{eqn:multiloss}) with gradients computed by automatic differentiation, it was noted that the optimization tended to add a constant vertical flux, $\overline{v} := (1/2\pi)\int_0^{2\pi} v \; dx$, to the solution.
The Navier-Stokes equations conserve $\overline{v}$, and changing it fundamentally changes the problem and the simple invariant solutions which we seek.

However, similar to \cite{Page2024} we find that the optimization is aided by allowing the vertical flux to vary from zero in the gradient descent.
This is possible because the adjoint equations \emph{do not} conserve $\overline{v}^{\dagger}$, hence net vertical flux is introduced into the guess $(\mathbf u, \Delta T, \alpha)$ upon computation of the gradient (\ref{eqn:opt_initial}) 
The mechanism by which vertical flux is created can be identified explicitly in the adjoint equations (\ref{eqn:adjoint}):
integrating the vertical adjoint velocity equation across the box yields
\begin{equation}
    \partial_t \overline{v}^{\dagger} = \frac{1}{2 \pi}\int_0^{2\pi}\mathbf v^{\dagger} \cdot \partial_y \mathbf u \; dx.
\end{equation}
We plot the evolution of the net adjoint flux (which is independent of $y$) in figure \ref{fig:mean_v} in terms of the backwards time $t^{\dagger} := T- t$.

At the conclusion of the optimization homotopy can be used to connect the finite-$\overline{v}$ solutions back to the no-flux case $\overline{v} = 0$.
We instead remove the mean flow by including it in the Newton problem, where we seek roots of
\begin{equation}
    \mathbf F(\mathbf u, T, \alpha) = \mathbf f^T(\mathscr T^{\alpha} \mathbf u) - \mathbf u + \overline{\overline{\mathbf u}},
\end{equation}
where $\overline{\overline{\mathbf u}}$ is a constant field of mean velocity:
\begin{equation}
\overline{\overline{\mathbf u}} = \hat {\bf x} \frac{1}{L_xL_y}\iint u\,dxdy + \hat {\bf y}\frac{1}{L_xL_y}\iint v\,dxdt\,. 
\end{equation}
This term was included in all of the Newton based optimisations reported in this letter: RPOs, tori, and tiles with turbulence.

\section{Further information concerning tiled solutions}
\label{sec:tables}
\begin{figure}
\centering
\includegraphics[width=6.3in]{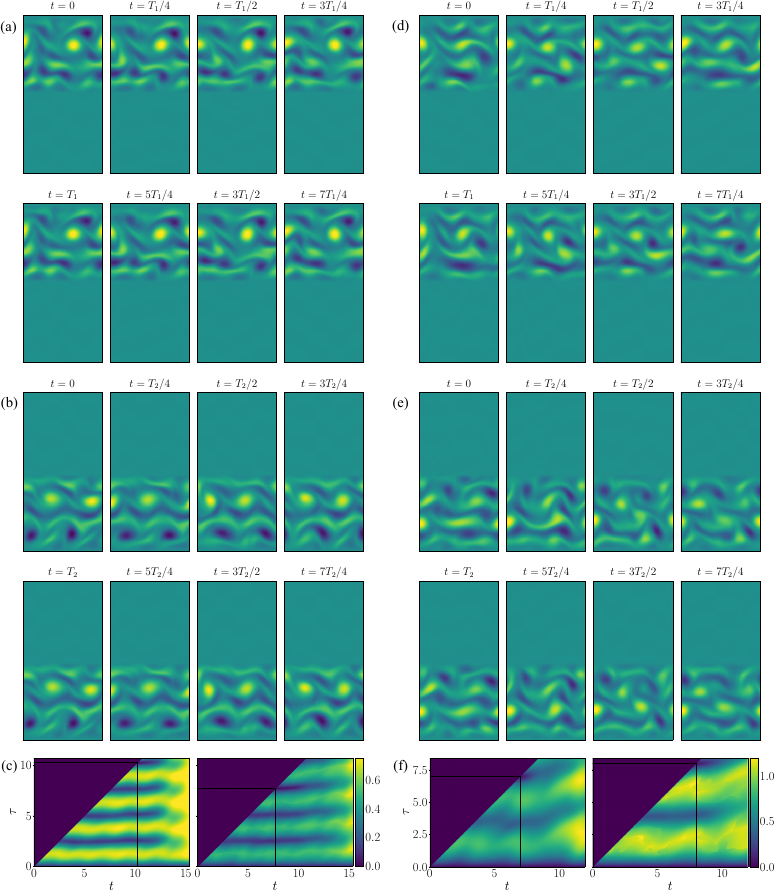}
\caption{\label{fig:rpo_SI} Trajectories of dynamically relevant two-tori constructed from pairs of small-box RPOs. Shown are contours of the out-of-plane vorticity $-11.4\leq \omega \leq 11.4$. (a) and (b) show the evolution of a two-torus constructed from RPOs 3 and 5 (see \S\ref{sec:tables}) -- each of (a) and (b) have been masked to isolate the dynamics in the two halves of the domain. Panels (d) and (e) show the same thing for the tiling of RPOs 4 and 5. All snapshots have been shifted in $x$ by $t_j \alpha_1/T_j$ to enable comparison between panels (note horizontally-aligned in the rows of (a) and (b) are separated by exactly one period). (c) and (f) show the masked autorecurrences $R(t, \tau)$ (see main paper) for (ab) and (de), respectively. 
}
\end{figure}

\begin{figure}
\centering
\includegraphics[width=3in]{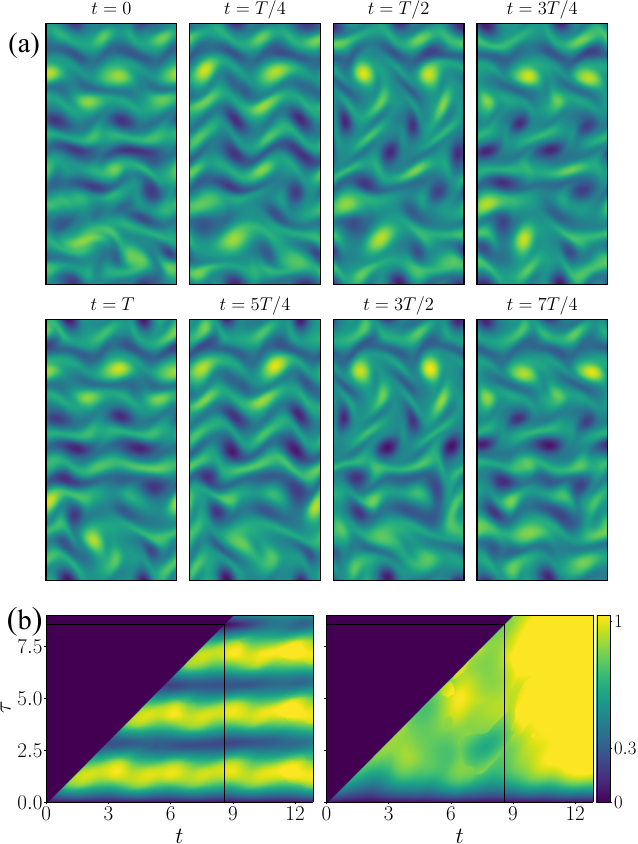}
\includegraphics[width=3in]{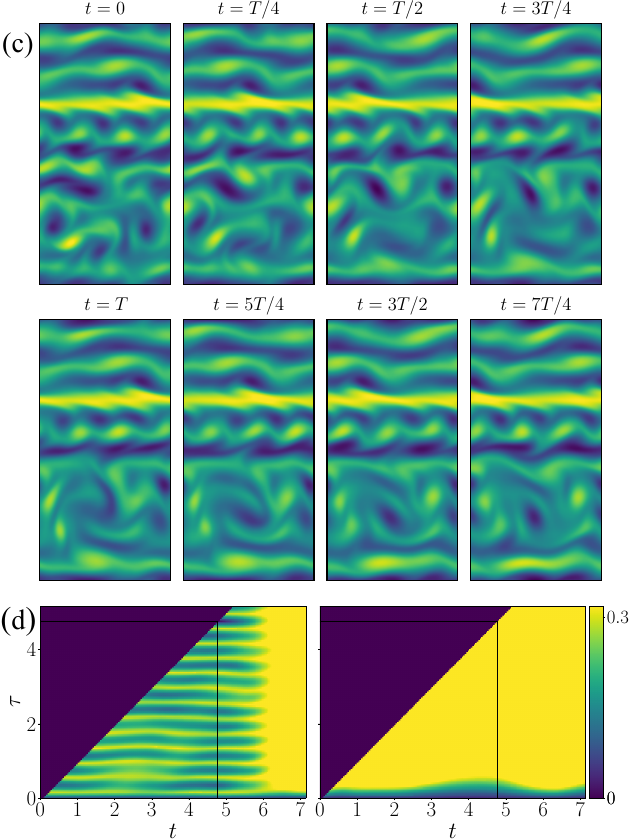}
\caption{\label{fig:turb_tile_SI} Two additional examples of RPO-turbulence tiles (ac) and their respective masked autorecurrences (bd). These solutions correspond to RPO numbers 3 and 16, respectively. Panels (a) and (c) show contours of the out-of-plane vorticity in the range $-11.4\leq \omega \leq 11.4$ .}
\end{figure}

\begin{figure}
\centering
\includegraphics[width=3.25in]{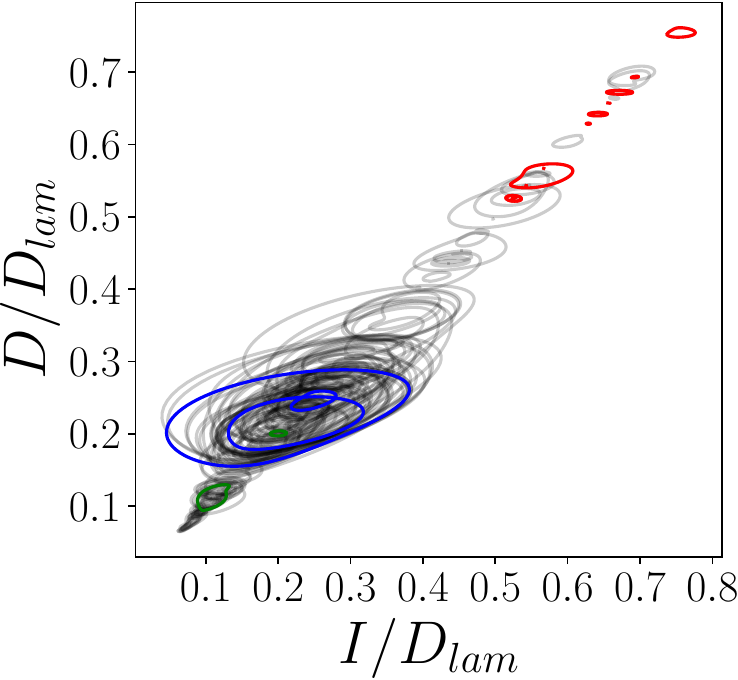}
\caption{\label{fig:PD_SI} The production dissipation plot of our original repository of $2\pi\times2\pi$ solutions (grey), the high dissipation RPOs that tiled (red), the low dissipation RPOs that tiled (blue), and additional low dissipation RPOs (green) that tiled exactly with non-laminar equilibria. }
\end{figure}

We include here further examples of time-series of trajectories shadowing two-tori (see figure \ref{fig:rpo_SI}) and turbulent trajectories shadowing small-box RPOs (see figure \ref{fig:turb_tile_SI}).
We also include a production-disspation plot (figure \ref{fig:PD_SI}) identifying which RPOs from the original repository (see \cite{Cleary2025}) were successfully used to generate new solutions in the large domain.

Finally, tables \ref{tab:lam_tiles}, \ref{tab:rpo_rpo_tiles} and \ref{tab:turb_tiles} summarize all the simple invariant solutions and more complex trajectories assembled in this work. 
In these tables, the `symmetry number' assigned to each RPO indicate the number of shift-reflect operations applied to the original RPO prior to tiling.



\begin{table}
    \centering
\begin{tabular}{c|c|c|c|c|c|c|c}
     RPO number & $T$&$T_{2\pi}$ &  $\alpha$&$\alpha_{2\pi}$&$D/D_{l}$ & $D_{2\pi}/D_{2\pi,l}$ & Symmetry number  \\
     \hline
     6 & 2.96 & 2.99 & 0.94 & 0.32 & 0.77 & 0.52 & 2\\
     6 & 2.96 & 2.99 & -0.94 & 0.32 & 0.77 & 0.52 & 4\\
     7 & 2.48 & 2.49 & 0.20 & 0.43 & 0.77 & 0.52 & 2\\
     7 & 2.46 & 2.49 & 0.09 & 0.43 & 0.77 & 0.52 & 7\\
     8 & 0.98 & 1.11 & 6.28 & 0.00 & 0.70 & 0.53 & 2\\
     8 & 0.98 & 1.11 & -0.00 & 0.00 & 0.70 & 0.53 & 5\\
     10 & 2.46 & 2.51 & -0.09 & 0.44 & 0.77 & 0.53 & 2\\
     10 & 2.48 & 2.51 & -0.20 & 0.44 & 0.77 & 0.53 & 7\\
     11 & 2.97 & 3.01 & -0.60 & -0.31 & 0.77 & 0.53 & 1\\
     12 & 1.97 & 2.01 & -0.77 & -1.19 & 0.77 & 0.53 & 1\\
     12 & 1.98 & 2.01 & -1.10 & -1.19 & 0.77 & 0.53 & 7\\
     13 & 2.44 & 2.44 & 5.89 & 0.26 & 0.77 & 0.54 & 3\\
     13 & 2.43 & 2.44 & 0.38 & 0.26 & 0.77 & 0.54 & 4\\
     14 & 3.90 & 3.90 & -1.06 & -2.11 & 0.77 & 0.54 & 2\\
     14 & 3.91 & 3.90 & -3.15 & -2.11 & 0.77 & 0.54 & 4\\
     15 & 3.75 & 3.17 & -2.67 & -0.09 & 0.79 & 0.56 & 2\\
     16 & 2.69 & 2.64 & -0.20 & 0.50 & 0.78 & 0.57 & 0\\
     17 & 1.75 & 2.00 & -0.15 & 0.51 & 0.82 & 0.63 & 0\\
     17 & 1.73 & 2.00 & 1.53 & 0.51 & 0.81 & 0.63 & 2\\
     17 & 2.11 & 2.00 & 1.92 & 0.51 & 0.90 & 0.63 & 4\\
     17 & 2.49 & 2.00 & -0.18 & 0.51 & 0.83 & 0.63 & 7\\
     18 & 1.39 & 2.47 & -4.06 & -0.82 & 0.85 & 0.64 & 7\\
     19 & 1.96 & 2.05 & 1.59 & 0.22 & 0.84 & 0.66 & 1\\
     20 & 3.20 & 4.35 & 0.65 & -1.02 & 0.83 & 0.67 & 6\\
     21 & 3.63 & 4.34 & 0.00 & -0.11 & 0.88 & 0.75 & 5\\
\end{tabular}

    \caption{A table of all of the successful tiles with the laminar flow. $T$ is the period of the converged solution, and $T_{2\pi}$ is the period of the initial solutions. The symmetry translation $\alpha$ and the laminar normalised mean dissipation $D/D_l$ .}
    \label{tab:lam_tiles}
\end{table}

\begin{table}
    \centering
    \begin{tabular}{c|c|c|c|c|c|c|c|c|c}
         RPO number 1 & RPO number 2 & $T_1$ & $T_2$& $n_1$ & $n_2$ & $D/D_l$ & Symmetry number 1 & Symmetry number 2  \\
         \hline
         6 & 16 & 1.50 & 2.64 & 6 & 3 & 0.55 & 4 & 0\\
         6 & 16 & 1.50 & 2.64 & 6 & 3 & 0.55 & 2 & 0\\
         19 & 6 & 2.05 & 1.50 & 3 & 4 & 0.60 & 1 & 4\\
         19 & 6 & 2.05 & 1.50 & 3 & 4 & 0.60 & 1 & 2\\
         14 & 16 & 1.95 & 2.64 & 4 & 3 & 0.55 & 4 & 0\\
         14 & 16 & 1.95 & 2.64 & 4 & 2 & 0.56 & 2 & 0\\
         3 & 4 & 2.90 & 2.98 & 2 & 3 & 0.21 & 0 & 0\\
         5 & 4 & 2.90 & 2.98 & 3 & 2 & 0.23 & 0 & 0\\
         1 & 4 & 3.00 & 2.98 & 4 & 3 & 0.21 & 0 & 0\\
         5 & 3 & 2.90 & 2.90 & 3 & 4 & 0.26 & 0 & 0\\
         2 & 5 & 2.92 & 2.90 & 2 & 3 & 0.26 & 0 & 0
    \end{tabular}
    \caption{Summary of all the candidate two-tori which are combinations of small-box RPO tiles. Here, $T_1$ and $T_2$ denote the periods of the original $2\pi\times2\pi$ solutions, and $n_1$ and $n_2$ are the number of periods for which we were able to successfully reduce the relative error to $\leq 0.01$. }
    \label{tab:rpo_rpo_tiles}
\end{table}

\begin{table}
    \centering
\begin{tabular}{c|c|c|c|c|c}
     RPO number 1 & $T$ & n & $\alpha$ & $D/D_l$ & Symmetry number 1   \\
     \hline
     3 &  3.36 &  2&2.60 & 0.19 & 2\\
     3 &  4.28 &  3&-3.82 & 0.20 & 6\\
     3 &  4.31 &  3&-5.02 & 0.21 & 6\\
     4 &  3.41 &  2&0.98 & 0.18 & 2\\
     4 &  2.70 &  2&0.62 & 0.24 & 2\\
     4 &  4.91 &  3&4.36 & 0.21 & 6\\
     5 &  4.09 &  3&0.02 & 0.22 & 6\\
     6 &  1.31 &  2&0.58 & 0.41 & 2\\
     6 &  1.47 &  2&-1.76 & 0.40 & 5\\
     6 &  2.24 &  3&0.80 & 0.46 & 6\\
     7 &  2.57 &  2&0.92 & 0.33 & 2\\
     8 &  2.20 &  2&1.21 & 0.40 & 2\\
     9 &  2.16 &  2&2.68 & 0.37 & 2\\
     10 &  2.59 &  2&0.92 & 0.33 & 2\\
     11 &  1.53 &  2&-1.99 & 0.44 & 2\\
     11 &  1.54 &  2&1.64 & 0.43 & 5\\
     11 &  1.32 &  2&1.44 & 0.45 & 5\\
     12 &  1.95 &  2&-2.40 & 0.42 & 2\\
     13 &  2.32 &  2&1.53 & 0.41 & 2\\
     13 &  2.44 &  2&-2.93 & 0.37 & 5\\
     14 &  1.80 &  2&1.50 & 0.38 & 2\\
     14 &  1.84 &  2&2.05 & 0.44 & 4\\
     14 &  1.84 &  2&2.42 & 0.50 & 4\\
     14 &  1.86 &  2&2.98 & 0.39 & 4\\
     14 &  1.83 &  2&-1.82 & 0.44 & 4\\
     14 &  1.65 &  2&-1.19 & 0.41 & 4\\
     14 &  1.71 &  2&-0.76 & 0.44 & 5\\
     14 &  1.77 &  2&-1.46 & 0.43 & 5\\
     16 &  2.38 &  2&0.21 & 0.41 & 0\\
     17 &  1.76 &  2&0.91 & 0.41 & 2\\
     19 &  1.02 &  2&-0.82 & 0.46 & 5\\
     19 &  0.96 &  2&-0.39 & 0.46 & 5\\
\end{tabular}
    \caption{A table of all of the successful RPO-turbulence tiles with relative error $\varepsilon \leq 0.01$ for at least two periods. in this case $T$ is the period of integration, and $\alpha$ is the symmetry shift. $D/D_l$ is the mean dissipation over a period after integrating out a transient.
    }
    \label{tab:turb_tiles}
\end{table}

\bibliography{refs.bib}

%% file: macros.tex
\newcommand{\svee}{{\hbox{\raise .4ex \hbox{${\scriptscriptstyle{\vee}}$}}}}
\newcommand{\swedge}{{\hbox{\raise .4ex \hbox{${\scriptscriptstyle{\wedge}}$}}}}

\newcommand{\CHI}{\hbox{\raise .4ex \hbox{$\chi$}}}

\newcommand{\scap}{\hbox{\raise .25ex
\hbox{${\operatornamewithlimits{\scriptstyle\bigcap}}$}}}
\newcommand{\scup}{\hbox{\raise .25ex
\hbox{${\operatornamewithlimits{\scriptstyle\bigcup}}$}}}

\newcommand{\deltacheck}
  {\overset {\lower .4ex \hbox{${\scriptscriptstyle{\hskip 2 pt\vee}}$}} \delta}

\newcommand{\Fcheck}
    {\overset {\lower .4ex \hbox{${\scriptscriptstyle{\hskip 2 pt\vee}}$}} F}

\newcommand{\fwedgehat}
    {\overset {\lower .6ex \hbox{${\scriptscriptstyle{\hskip 3 pt\wedge}}$}} f}
\newcommand{\fveecheck}
    {\overset {\lower .4ex \hbox{${\scriptscriptstyle{\hskip 2 pt\vee}}$}} f}
\newcommand{\fkcheck}
   {\overset {\lower .4ex \hbox{${\scriptscriptstyle{\hskip 1 pt\vee}}$}} {f_k}}

\newcommand{\raiseprime}{\hbox{\raise .3ex \hbox{${\scriptstyle{\prime}}$}}}

\newcommand{\Gcheck}
    {\overset {\lower .4ex \hbox{${\scriptscriptstyle{\hskip 2 pt\vee}}$}} G}

\newcommand{\gveecheck}
    {\overset {\lower .4ex \hbox{${\scriptscriptstyle{\hskip 2 pt\vee}}$}} g}

\newcommand{\hcheck}
    {\overset {\lower .4ex \hbox{${\scriptscriptstyle{\hskip 2 pt\vee}}$}} h}

\newcommand{\Kcheck}
  {\overset {\lower .4ex \hbox{${\scriptscriptstyle{\hskip 2 pt\vee}}$}} K}

\newcommand{\mucheck}
    {\overset {\lower .4ex \hbox{${\scriptscriptstyle{\hskip 2 pt\vee}}$}} \mu}

\newcommand{\norm}[1]{\|#1\|}

\newcommand{\varphicheck}
    {\overset {\lower .4ex \hbox{${\scriptscriptstyle{\hskip 1 pt\vee}}$}}
              \varphi}

\newcommand{\Rbar}
  {{\overset {\hskip -0.9 pt \lower 1.5pt \hbox{{\rule{6.7pt}{0.45pt}}}} \R}}
\newcommand{\subRbar}
   {{\overset {\hskip -0.8 pt \lower 1.5pt \hbox{{\rule{4.5pt}{0.5pt}}}} \R}}

\newcommand{\zeroveecheck}
    {\overset {\lower .4ex \hbox{${\scriptscriptstyle{\hskip 0.5 pt\vee}}$}} 0}

\newcommand{\eb}{\begin{equation}\begin{aligned}}
\newcommand{\ub}{{\bf u}}

\newcommand{\en}{\end{aligned}\end{equation}}
\newcommand{\mnorm}[1]{{\left\vert\kern-0.25ex\left\vert\kern-0.25ex\left\vert #1 \right\vert\kern-0.25ex\right\vert\kern-0.25ex\right\vert}}